\documentclass[twocolumn,showpacs,PRB]{revtex4}

\usepackage{bm}
\usepackage{amssymb}
\usepackage{amsmath}
\usepackage{pifont}
\usepackage{graphics}
\usepackage[sort&compress]{natbib}
\usepackage{epsfig}
\usepackage{color}

\usepackage[ps2pdf,colorlinks=true, pagebackref=false, bookmarks=true,bookmarksopen=true,bookmarksnumbered=true]{hyperref}

\newcommand{\Cu}{\ensuremath{\mathrm{Cu}}}
\newcommand{\Ni}{\ensuremath{\mathrm{Ni}}}
\newcommand{\Nb}{\ensuremath{\mathrm{Nb}}}
\newcommand{\Al}{\ensuremath{\mathrm{Al}}}

\renewcommand{\O}{\ensuremath{\mathrm{O}}}
\newcommand{\Co}{\ensuremath{\mathrm{Co}}}
\newcommand{\Fe}{\ensuremath{\mathrm{Fe}}}
\newcommand{\Pd}{\ensuremath{\mathrm{Pd}}}

\newcommand{\W}[1]{{#1}}

\begin{document}


\title{%
 Josephson tunnel junctions with strong ferromagnetic interlayer
}

\author{A. A. Bannykh}

\affiliation{%
   Institute of Solid State Research and JARA- Fundamentals of Future Information Technology, Research Centre, J\"ulich, 52425 J\"ulich, Germany %
}
\affiliation{Institute of Solid State Physics, Russian Academy of Sciences, Chernogolovka, 142432, Russia}

\author{J. Pfeiffer}
\affiliation{%
  Physikalisches Institut-Experimentalphysik II and Center for Collective Quantum Phenomena,   Universit\"at T\"ubingen,
  Auf der Morgenstelle 14,
  72076 T\"ubingen, Germany
}
\author{V. S. Stolyarov}
\affiliation{Institute of Solid State Physics, Russian Academy of Sciences, Chernogolovka, 142432, Russia}

\author{I. E. Batov}
\affiliation{Institute of Solid State Physics, Russian Academy of Sciences, Chernogolovka, 142432, Russia}

\author{V. V. Ryazanov}
\affiliation{Institute of Solid State Physics, Russian Academy of Sciences, Chernogolovka, 142432, Russia}

\author{M. Weides}
\email{m.weides@fz-juelich.de}
\affiliation{%
   Institute of Solid State Research and JARA- Fundamentals of Future Information Technology, Research Centre, J\"ulich, 52425 J\"ulich, Germany%
   }

\date{\today}

\begin{abstract}
The dependence of the critical current density $j_c$  on the ferromagnetic interlayer thickness $d_F$ was determined for $\Nb/\Al_2\O_3/\Cu/\Ni/\Nb$ Josephson tunnel junctions with ferromagnetic $\Ni$ interlayer thicknesses from very thin films ($\sim1\:\rm{nm}$) upwards and classified into F-layer thickness regimes showing a dead magnetic layer, exchange, exchange + anisotropy and total suppression of $j_c$. The Josephson coupling changes from $0$ to $\pi$ as function of $d_F$, and -very close to the crossover thickness- as function of temperature. The strong suppression of the supercurrent in comparison to non-magnetic $\Nb/\Al_2\O_3/\Cu/\Nb$ junctions indicated that the insertion of a F-layer leads to additional interface scattering. The transport inside the dead magnetic layer was in dirty limit. For the magnetically active regime fitting with both the clean and the dirty limit theory were carried out, indicating dirty limit condition, too. \W{The results were discussed in the framework of literature}.
\end{abstract}



\maketitle

\section{Introduction}

The combination of superconducting (S) and ferromagnetic (F) materials in layered structures leads to phase oscillations of the superconducting wave-function inside the ferromagnet \cite{buzdin05RMP}. If the F-layer thickness $d_F$ in SFS Josephson junctions (JJs) is of the order of one half of this oscillation wave length, the wave function changes its sign, i.e., shifts its phase by $\pi$ while crossing the F-layer. In this case the critical current $I_c$ (and critical current density $j_c$) turns out to be negative and the current-phase relation reads $I = I_c \sin(\phi) = |I_c| \sin(\phi+\pi)$ with $I_c < 0$, $\phi$ being the phase difference between the two superconducting electrodes. Such JJs are called $\pi$ JJs because their phase difference is $\phi +\pi$ in the ground state \cite{buzdin05RMP}. Conventional JJs are called $0$ JJs because they have a current-phase relation of $I = I_c \sin(\phi)$ with $I_c > 0$ and the ground phase difference $\phi = 0$. The insertion of an insulating barrier I in SFS stacks, i.e., SIFS stacks, is advantageous as the damping of Josephson phase dynamics becomes lower and the voltage drop gets larger. \W{This facilitates both the study of dynamics and the transport measurements.}\\
The most convincing demonstration of the phase oscillations in SFS/SIFS structures is the \emph{damped oscillatory behavior} of the critical current $I_c$ in the F-layer as a function of temperature $T$ \cite{Ryazanov01piSFS_PRL,Sellier03TinducedSFS} or of the F-layer thickness $d_F$ \cite{Kontos02Negativecoupling,OboznovRyazanov06IcdF,WeidesHighQualityJJ}. The decay length is $\xi_{F1}$ and the oscillation length $2\pi\xi_{F2}$. $\xi_{F1}$ is based on the well-known proximity effect, i.e., the exponential decay of the Cooper pair density inside a metal adjacent to a superconductor. A quantitative model in the \emph{dirty limit} where the mean free path $\ell<d_F,\:\hbar v_F/E_\mathrm{ex}$, with $v_F$ being the Fermi velocity, $E_\mathrm{ex}$ being the magnetic exchange energy, can be found in Ref. \onlinecite{VasenkoPRB}. This model utilizes parameters which characterize the material properties of the S and F-layers and the S/F interface transparency. At $T \lesssim T_c\ll E_\mathrm{ex}/k_B$
\begin{equation}
  I_{c}(d_F) \sim
  \exp\left( \frac{-d_F}{\xi_{F1}} \right)
  \cos \left( \frac{d_F-d_F^\mathrm{dead}}{\xi_{F2}}\right)
  ,\label{Eq:IcRn}
\end{equation}
where%
\begin{align*}\xi_{F1,F2}&=\sqrt{\frac{\hbar v_F\ell}{3E_\mathrm{ex}}}\left({\sqrt{\sqrt{1+\underbrace{\left(\frac{\hbar}{E_\mathrm{ex}\tau_m}\right)^2}_{\rightarrow0}}\pm\frac{\hbar}{E_\mathrm{ex}\tau_m}}}\right)^{-1}\\
&\approx\sqrt{\frac{\hbar v_F \ell}{3E_\mathrm{ex}}}\left({1\pm \frac{\hbar}{2 E_\mathrm{ex}\tau_m}}\right)^{-1}\;.
\end{align*}
The parameter $\hbar/E_\mathrm{ex}\tau_m$ is considered as being much smaller than unity. $\tau_m$ is the inelastic magnetic scattering time and $d_F^\mathrm{dead}$ the magnetic dead layer thickness. Within the framework of this theory $\xi_{F1}$ is shorter than $\xi_{F2}$. Strictly speaking, Eq. \ref{Eq:IcRn} is only valid close to $T_c\approx9\:\rm{K}$ for $\Nb$-based JJs, whereas most of our samples were measured at $4.2\:\rm{K}$. However, the $I_c(d_F,T)$ dependence with an arbitrary temperature $T$ is a complex sum depending on Matsubara frequencies, spin-flip scattering time and exchange energy. It was shown in Ref. \cite{OboznovRyazanov06IcdF} that the corrections originating from this complex approach are important only for calculations of $I_c(T)$ dependencies close to the $0$ to $\pi$ transition points. The approach used in Eq. \ref{Eq:IcRn} yields good results being suitable for fitting of $I_c(d_F)$ dependencies in a wide temperature range. \\%
Experimentally for JJs in dirty limit \cite{OboznovRyazanov06IcdF,WeidesHighQualityJJ} not more than two oscillations of $I_c(d_F)$ were observed. Below $T_c$ the temperature variation  of $E_\mathrm{ex}$ is negligible for $E_\mathrm{ex}\gg k_B T$ and $I_c(T)$ depends mostly on the temperature-sensitive effective magnetic scattering rate. For example, temperature-driven changes of the coupling were observed in Ref. \cite{Sellier03TinducedSFS,OboznovRyazanov06IcdF,WeidesHighQualityJJ}.\\
Contrary, in the \emph{clean limit}, where $\ell>d_F$, the simple clean limit theory for $T$ near $T_c$ (Ref. \cite{buzdin05RMP}) is
\begin{equation}
  I_{c}(d_F) \sim \frac{\sin\left(\frac{2E_\mathrm{ex} }{\hbar v_F}(d_F-d_F^\mathrm{dead})\right)}{\frac{2E_\mathrm{ex} }{\hbar v_F}(d_F-d_F^\mathrm{dead})}.\label{Eq:IcRn_clean}
\end{equation}
The decay of $I_c\sim1/d_F$  can be much slower than its oscillation period. As consequence, several phase oscillations may be experimentally detectable. For example, multiple $I_c(d_F)$ oscillations were possibly observed in SFS-JJs using elemental magnets such as $\Ni$ \cite{Blum:2002:IcOscillations,Blum04NanoGiga,ShelukhinSFS06,RobinsonPRB07}. Unfortunately, insufficient density of data points has not allowed to carry out a reliable quantitative comparison between theory and experiment. Although the absolute $I_c(d_F)$ dependencies for all sets \cite{Blum:2002:IcOscillations,Blum04NanoGiga,ShelukhinSFS06,RobinsonPRB07} of \emph{clean} $\Ni$-SFS junctions are hardly comparable, as a general feature the decay of adjacent maxima amplitudes is below a factor of $4\textrm{-}5$, much smaller than the observed factor of $10^4$ as it is the case with \emph{dirty} $\Ni\Cu$-SFS junctions \cite{OboznovRyazanov06IcdF}. SIFS stacks in the clean limit may be used to obtain a high critical current density $j_c$ in the $\pi$ state, which -for example- is advantageous to obtain $0$--$\pi$ junctions  \cite{WeidesFractVortex,WeidesSteppedJJ} in the long Josephson limit and to study the dynamics of fractional vortices \cite{PfeifferPRB08}.\\
The clean limit theory from Ref.\cite{buzdin05RMP}, i.e., Eq. \ref{Eq:IcRn_clean}, yields no temperature driven $0$ to $\pi$ transition for $E_\mathrm{ex}\gg k_BT$ and temperature independent $E_\mathrm{ex}$. A more complex theory \cite{RadovicPRB03} for strong magnets (like $\Ni$) and insulating interfaces, i.e., tunnel barriers, predicts that samples with $d_F$ very close to the crossover thickness $d_F^{0\textrm{-}\pi}$ may change their ground state with temperature. \W{However, up to now, a temperature driven phase transition for JJs in the clean limit or for SFS-type JJs with elemental magnetic interlayer was not reported, yet.}

\W{Diluted magnetic alloys ($\Cu\Ni$, $\Pd\Ni$) contain numerous spin-flip centers (e.g. $\Ni$-rich clusters) that increase superconducting order parameter decay. So the elemental magnet use in SIFS junction can yield some advantages.} In this paper, we study $\Ni$-SIFS junctions starting with $d_F\sim1\:\rm{nm}$ upwards and a high density of data points along the $j_c(d_F)$ dependence. In particular, we show details of the Fraunhofer pattern $I_c(H)$ of these junctions and their $I_c(T)$ dependence for various thicknesses. The results are divided into two parts: Section \ref{Sec:experiment} addresses the fabrication, $j_c(d_F)$, $I_c(T)$ and $I_c(H)$ measurements and in Section \ref{Sec:discussion} we discuss our results using the transport theories in clean and dirty limit \W{and compare it with literature}.

\section{Experiment}\label{Sec:experiment}

\begin{figure}[tb]
\includegraphics[width=8.6cm]{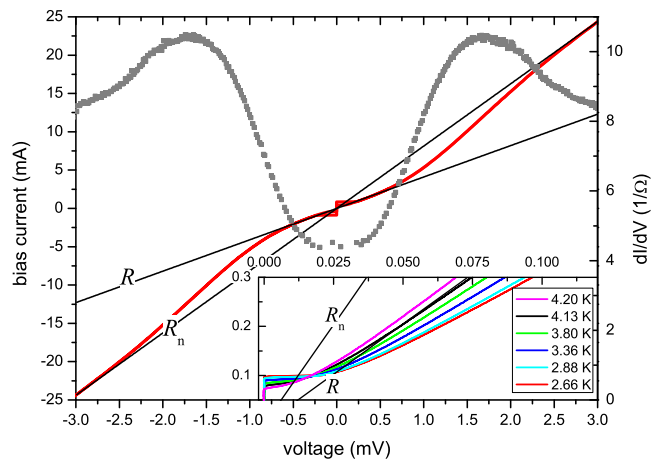}
  \caption{\W{(Color online) (a) IV and dI/dV characteristics (square symbols) at $4.2\:\rm{K}$ of $d_F=1.4\:\rm{nm}$ sample. The inset depicts IVC for $d_F=2.6\:\rm{nm}$. Both the normal and subgap resistances $R_n$ and $R$ are plotted (gray lines).}}\label{IVC}
\end{figure}

\begin{figure}[tb]
\includegraphics[width=8.6cm]{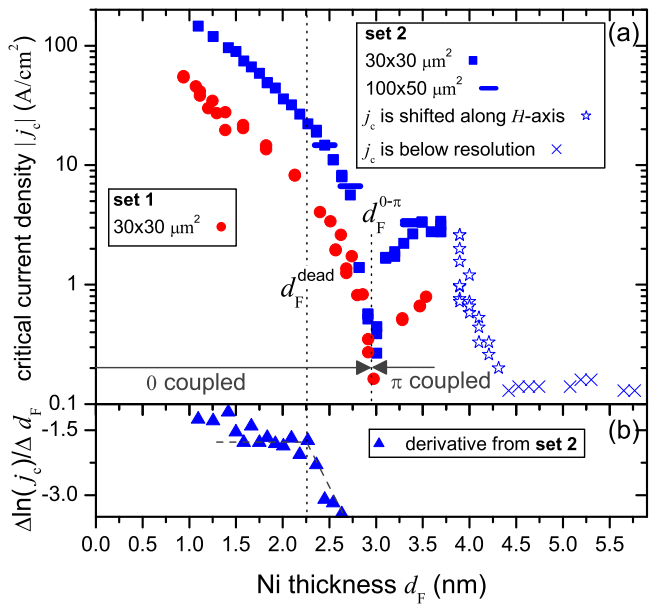}
  \caption{(Color online) (a) $j_c(d_F)$ dependence for two sets of SIFS-JJs. Set 1 has a thicker $\Al_2\O_3$ barrier than set 2. Standard $I_c(H)$ pattern were obtained, except for i) samples with \ding{73} having strongly shifted $I_c(H)$ pattern (see Ref. \cite{WeidesAnisotropySIFS}) and ii)  samples with thick $d_F$ showing no $I_c(H)$ pattern, i.e., $I_c$ is below the measurement resolution $\times$. The coupling changes from $0$ to $\pi$ at $d_F^{0\textrm{-}\pi}=2.95\:\rm{nm}$. Measurements were taken at $4.2\:\rm{K}$. (b) The dead magnetic layer regime is estimated by the change in slope of derivative of $\ln j_c(d_F)$ as $d_F^\mathrm{dead}=2.26\:\rm{nm}$. The dashed lines in (b) are guide to the eyes.}\label{IcdF}
\end{figure}
To produce high quality SIFS-JJs one has to control both thickness and interface roughness of the F-layer on a sub-nm scale. The multilayers were \W{computer-controlled} sputter deposited at room-temperature at a background pressure of $5\cdot10^{-7}\:\rm{mbar}$ on $4\:\rm{Inch}$ wafers. The uniform growth of the $\Ni$ layer was ensured by a thin $2\:\rm{nm}$ $\Cu$ interlayer between the I-layer (necessarily having flat interfaces) and the F-layer \cite{WeidesFabricationJJPhysicaC}. Thus, the stack was actually SINFS-type like. \W{The presence of $\Cu$ does not influence much the IVC as determined by SIS and SINS-type junctions, due to the strong proximity effect of $\Cu$}. The $\Nb$ electrodes had thicknesses of $150\;\rm{nm}$ (bottom) and $400\;\rm{nm}$ (top). \W{Anodic oxidation spectroscopy on reference SIS samples, XRD and profiler measurements of the sputter rates and specific resistance measurement of $\Nb$ thin films have been made to control the quality of films. The $\Nb$ bottom electrode was made up by four $37\:\rm{nm}$ Nb layers, each separated by $2.4\:\rm{nm}$ $\Al$ layers to reduce the total roughness \cite{kohlstedtNbALNb}.} One wafer contained JJs with different $d_F$ deposited in a single run by shifting the substrate and the $\Ni$-sputter target \cite{WeidesFabricationJJPhysicaC}. The estimation of the F-layer thickness $d_F$ yields values which do not reflect the finite diameter of $\Ni$ atoms, but the polycrystalline growth of F-layer may thoroughly permit a steadily change of the \emph{effective} F-layer thickness by delicate variation of the sputtering rate. \W{A systematic, absolute error in $d_F$ due to an off-centered wafer during deposition is minimized by a special wafer clamp. The relative error due to an not-ideal wedge-shaped F-layer (having two gradients, one parallel, and a much smaller one perpendicular to the wafer axis) was minimized by taking JJs being located maximally ca. $1\:\rm{mm}$ apart the wafer axis. We estimate the relative error in $d_F$ as less than 5\%.} The $\Al_2\O_3$ tunnel barrier was formed statically ($5\;\rm{mbar}$ partial oxygen pressure, set 1) or dynamically ($0.017\;\rm{mbar}$, set 2) for $30\;\rm{min}$ at room temperature. The JJs had areas of $30\times 30$ and $100\times 50\:\mathrm{\mu m^2}$. The lateral sizes of these junctions were comparable or smaller than the Josephson penetration length $\lambda_J$. \W{The transport measurement, i.e. $I_c(H)$ and IVC, gave no indication of the existence of a superconducting short either inside the $\Al_2\O_3$ tunnel barrier nor in the insulating $\Nb_2\O_5$ barrier.}\\
The samples were cooled down using $\mu$-metal or cryoperm shields to suppress stray fields. Transport measurements were performed in liquid He either using a dip stick setup or a cryostat with the respective inset. The cryostat could reach temperatures between $1.3\textrm{-}10\:\rm{K}$. Standard room-temperature voltage amplifiers were used. A magnetic field $H$ was applied in-plane and parallel to the longer sample axis (regarding the $100\times 50\:\mathrm{\mu m^2}$ samples). \W{The current bias was computer-controlled statically swept while measuring the voltage drop across the junction. Both current and voltage values were automatically averaged over several hundreds of data points for each step. The upper limit of $I_c$ was determined by a given voltage criteria.}
\begin{figure}[bt]
\includegraphics[width=8.6cm]{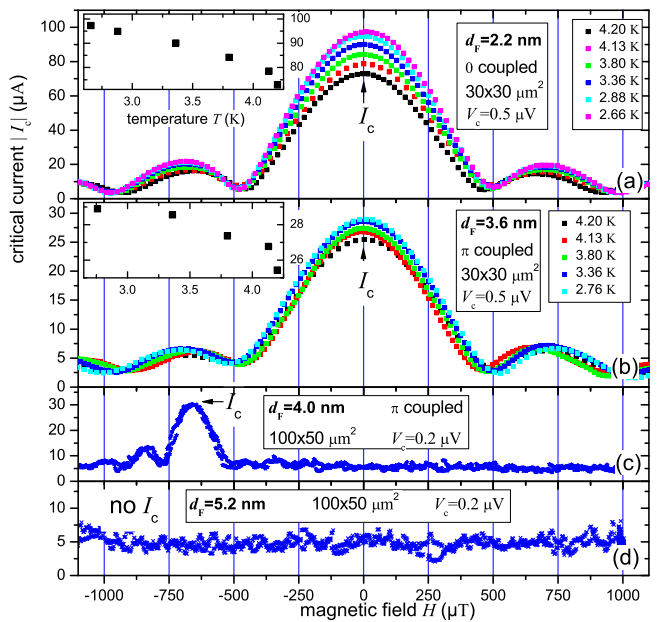}
  \caption{(Color online) $I_c(H)$ dependence of JJs from set 2 for different thicknesses $d_F$. $I_c(T)$ for samples from a) and b) are plotted in the inset. Magnetic field was applied along the long axis of the $100\times 50\:\mathrm{\mu m^2}$ samples. The strongly shifted $I_c(H)$ pattern from c) is discussed in Ref.~\cite{WeidesAnisotropySIFS}. c) and d) were measured at $4.2\:\rm{K}$.} \label{IcHT}
\end{figure}

\begin{figure}[t]
\includegraphics[width=8.6cm]{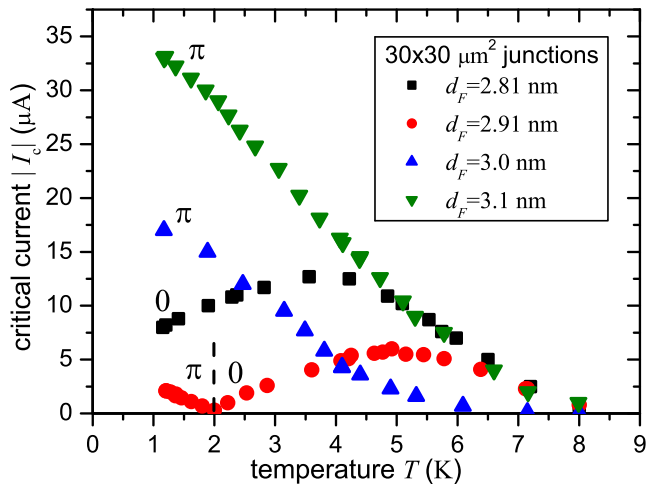}
  \caption{(Color online) $I_c(T)$ dependence for JJs from set 2 in vicinity of the crossover thickness $d_F^{0\textrm{-}\pi}$. A temperature-induced $0$ to $\pi$ transition is observed (sample $d_F=2.91\:\rm{nm}$). $d_F^{0\textrm{-}\pi}$ is shifted from $2.95\:\rm{nm}$ ($4.2\:\rm{K}$) to $2.91\:\rm{nm}$ ($2.0\:\rm{K}$).} \label{IcT}
\end{figure}

\W{The IV and dI/dV characteristics (square symbols) at $4.2\:\rm{K}$ of $d_F=1.4\:\rm{nm}$ sample together with the temperature dependence of IVC for $d_F=2.6\:\rm{nm}$ sample are plotted in Fig. \ref{IVC}. The data close to $0\:\rm{\mu V}$ is removed from the dI/dV graph. The maximum conductivity appears for voltages close to the superconducting gap ($1.7\:\rm{mV}$) of the bottom electrode. The gap of the top electrode is covered by the large subgap current in F-layer. Both the normal and subgap resistances $R_n$ and $R$ are plotted (gray lines).}\\
Fig. \ref{IcdF} depicts the  $j_c(d_F)$ dependencies for both sets of samples. Note the logarithmic scale of the $j_c$-axis. The generally larger $j_c$'s for set 2 reflect the thinner $\Al_2\O_3$ tunnel barrier than in set 1. Samples with $d_F\lesssim2.0\:\rm{nm}$ showed an underdamped behavior, i.e., a hysteretic IV curve, at $4.2\:\rm{K}$ (data not shown). Normal state and subgap resistance indicate a small variation ($\sim5\%$) for JJs with the same $d_F$.
All JJs up to $d_F=3.8\:\rm{nm}$ (solid symbols) had standard $I_c(H)$ pattern (see Fig. \ref{IcHT} a,b) and showed a small junction to junction variation of $j_c$ ($\sim5\%$). Between $d_F=3.8\textrm{-}4.4\:\rm{nm}$ (\ding{73} in Fig. \ref{IcdF}) the JJs had strongly shifted $I_c(H)$ pattern, as depicted in Fig. \ref{IcHT} c); the magnetic origin due to anisotropy effects was discussed in Ref. \onlinecite{WeidesAnisotropySIFS}. For the respective fitting procedure for each $d_F$ only the largest $I_c$'s, measured at finite $H$, were used. For $d_F>4.4\:\rm{nm}$ no $I_c(H)$ pattern could be measured, as no dependence on applied magnetic field was observed (Fig. \ref{IcHT} d) due to the suppression of $I_c$ below the measurement resolution. The upper limit of $I_c$ measured with voltage criteria $V_c=0.2\rm{-}0.5\:\rm{\mu V}$ is depicted by $\times$.\\
At $d_F^{0\textrm{-}\pi}=2.95\:\rm{nm}$ the Josephson phase changes from $0$ to $\pi$. There is no indication that another minimum occurs before, also not below $1\:\rm{nm}$, as this is inside the dead magnetic layer regime (see Sec. \ref{thicknessDiscussion}). The maximum $j_c$ in the $\pi$ state is $\sim3.4\:\rm{A/cm^2}$.\\
The $I_c(T)$ dependence of samples is shown in the vicinity and apart from the phase transition thickness  $d_F^{0\textrm{-}\pi}$ in Fig. \ref{IcT} and insets of Fig. \ref{IcHT}.

\section{Discussion}\label{Sec:discussion}
In this section we discuss the transport properties of our $\Ni$ film, the temperature induced phase transition and compare our $j_c(d_F)$ data with clean and dirty limit theory. For discussion we used only the data from set 2, as set 1 contains less data, and it's $j_c(d_F)$ dependence is similar to set 2, with lower amplitude of $j_c$.
\subsection{Mean free path}
The specific resistance $\rho=7.4\;\rm{\mu\Omega cm}$ at $4.2\;\rm{K}$ was determined by a four-point in-plane measurement of a $3.2\:\rm{nm}$ thin $\Ni$ film. However, one should consider the importance of grain boundary scattering on the in-plane resistance for thin films. In the JJs current transport is out-of-plane, where $\rho$ becomes smaller. The estimation of $\ell$ was problematic, as there is a considerable spread of data in the literature, depending strongly on the quality of $\Ni$ film. Assuming a constant Fermi surface and the Pippard relation \cite{Pippard60} $\ell=\pi^2 k_B^2/\gamma v_F e^2\rho$ with a bulk specific heat constant $\gamma$ yields $\ell=3.3\:\rm{nm}$, while considering the complex Fermi surface of $\Ni$ gives $\ell=0.7\textrm{-}2.3\cdot10^{-15}\:\rm{\Omega m^2}/\rho=9\textrm{-}30\:\rm{nm}$ \cite{Fierz90}. This estimation was confirmed by measurements of the out-of-plane length  $\ell\approx21\:\rm{nm}$ \cite{MoreauMFPNi} for samples with in-plane specific resistance of $\rho=3.3\;\rm{\mu\Omega cm}$, i.e., roughly half of our value. However, spin- and angle-resolved photoemission \cite{PetrovykhNI1998} on some other $\Ni$ samples yields a spin independent, very short mean free path $\ell\approx2\:\rm{nm}$.\\The dirty limit condition $\ell<d_F$ is not valid over the total range of $d_F=1\textrm{-}6\:\rm{nm}$ if considering the smaller values of $\ell$, for larger values of $\ell$ it is not valid at all. To answer the question, whether our samples are in the clean or dirty limit,
is not that easy. Therefore in Section \ref{thicknessDiscussion} we compare the $j_c(d_F)$ dependence both with clean and dirty limit Eq.~\ref{Eq:IcRn_clean} and Eq.~\ref{Eq:IcRn}.

\subsection{Temperature dependence of $I_c$}
In Fig. \ref{IcT} the $I_c(T)$ dependence for four samples in the vicinity of the thickness-induced $0$ to $\pi$ transition is shown. By decreasing the temperature a $0$ to $\pi$ transition is observed for one sample ($d_F=2.91\:\rm{nm}$). The phase transition thickness $d_F^{0\textrm{-}\pi}$ varies from $2.95\:\rm{nm}$ at $4.2\:\rm{K}$ down to $2.91\:\rm{nm}$  at $2.0\:\rm{K}$.  To our knowledge, this is the first observation of a $T$-induced $0$ to $\pi$ transition for SIFS junctions using an elemental magnet. Temperature induced $0$ to $\pi$ transitions were observed in dirty SFS stacks having transparent SF-interfaces \cite{OboznovRyazanov06IcdF}, and theoretically predicted for clean SIFIS stacks \cite{RadovicPRB03}. For the dirty SIFS stacks,  having one transparent SF-interface, it was for the first time observed in $\Ni\Cu$-based JJs \cite{WeidesHighQualityJJ}. We are not aware of the experimental observation of temperature induced $0$ to $\pi$ transition in presumably clean SFS, SIFS, or SIFIS stacks. Thus it's occurrence in our $\Ni$-SIFS stacks is an indication for being in the dirty limit condition and having one transparent SF-interface.\\
The sample with $d_F=2.81\:\rm{nm}$ in Fig. \ref{IcT} shows an anomaly in $I_c(T)$ below $4.0\:\rm{K}$, but does not change the ground state. For JJs with $d_F$ over $0.2\:\rm{nm}$ from $d_F^{0\textrm{-}\pi}$ normal $I_c(T)$ dependencies were measured, see insets of Fig. \ref{IcHT}.\\
In principle, one may try to fit the general form of $I_c(d_F,T)$ to the measured $I_c(T)$ dependencies. However, among parameters like exchange energy $E_\mathrm{ex}$, Fermi velocity $v_F$, spin flip scattering time $\tau_m$ and thickness $d_F$ one should enter the mean free path $\ell$, which can not be determined precisely for our samples.

\subsection{F-layer thickness dependence of $j_c$}\label{thicknessDiscussion}
\begin{figure}[tb]
\includegraphics[width=8.6cm]{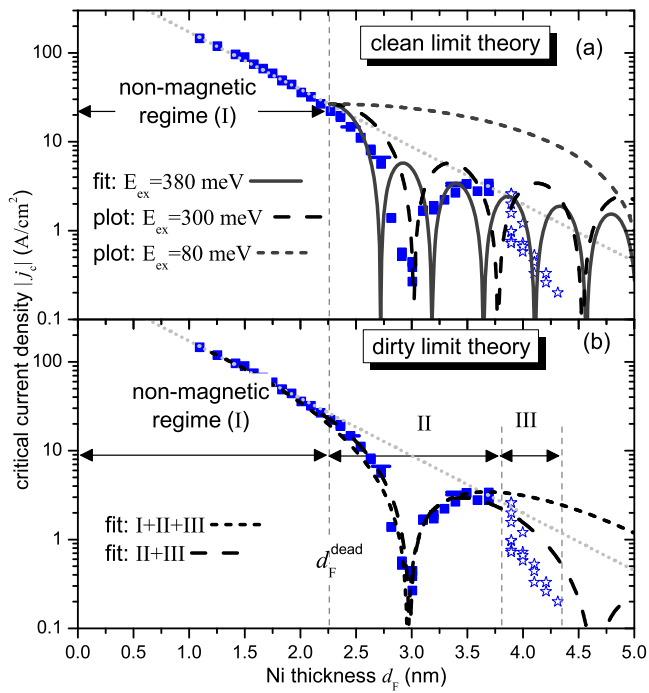}
  \caption{(Color online) $j_c(d_F)$ dependence for set 2. The data set is split into different regimes: non-magnetic (denoted by I), magnetic exchange (II) and magnetic exchange + anisotropy (III). The fit to regime I (dotted line) is extrapolated to the magnetically active regime. (a) The \emph{clean} limit dependence (Eq. \ref{Eq:IcRn_clean}) was fitted (solid line, yielding $E_\mathrm{ex}=380\:\rm{meV}$) and plotted (dashed lines) for $E_\mathrm{ex}=80$ and $300\:\rm{meV}$. Starting F-layer thickness is $d_F^\mathrm{dead}$ for all three curves. (b) Two fits for \emph{dirty} limit dependence (Eq. \ref{Eq:IcRn}) for data from regimes I+II+III and just II+III. For more details, see text.}\label{IcdFdirty}
\end{figure}

For smallest F-layer thicknesses $d_F$ the JJs are supposed to be inside the dead magnetic layer regime, and JJs with thicker $d_F$ should show a  $j_c(d_F)$ dependence being influenced by the exchange energy. We split the data into different F-layer thickness regimes showing a dead magnetic layer, exchange and exchange+anisotropy, denoted by I, II and III.

\subsubsection*{Non-magnetic interlayer regime}
Inside the dead magnetic layer $0\:\mathrm{nm}<d_F<d_F^\mathrm{dead}$ the wave-function amplitude is damped like in non-magnetic normal metal. A dead magnetic layer can be caused by polycrystalline growth, interdiffusion at both interfaces ($\Cu$ or $\Nb$) and the rather weak magnetic moment of $\Ni$. The local magnetic moments are uncoupled, and have random orientations, i.e., the material is paramagnetic. In the literature \cite{RobinsonPRB07} $d_F^\mathrm{dead}$ in SFS-JJs with F=$\Ni$ was determined as $1.3\:\rm{nm}$, indicating that its influence on the supercurrent transport is not negligible. In our case regime I ($d_F\leq d_F^\mathrm{dead}$) covers JJs which show the normal proximity effect due to leakage of Cooper pairs in the dead magnetic layer:
\begin{equation}
j_c =j_c^0\exp{\left(\frac{-d_F}{\xi_{F}^{\textrm{dead}}}\right)}
.\label{Eq:IcRn_dead}
\end{equation}
$j_c^0$ is the maximum amplitude without $\Ni$-layer, i.e., of a SINS junction.
$d_F^\mathrm{dead}$ is estimated as $2.26\:\rm{nm}$ by the change in slope of $\Delta \ln (j_c)/\Delta d_F$, see Fig. \ref{IcdF} b). The dashed lines serve as guide to the eyes. Fitting Eq. \ref{Eq:IcRn_dead} to regime I of set 2 yields $j_c^0=0.76\:\rm{kA/cm^2}$ and $\xi_{F}^{\textrm{dead}}=0.68\:\rm{nm}$, see dotted line in Fig. \ref{IcdFdirty}. The measured value for a SINS-type junction (N=$2\:\rm{nm}\;\Cu$) is $j_c^0=4\:\rm{kA/cm^2}$, indicating some additional scattering at the $\Ni$ interfaces of SINFS stack.

\subsubsection*{Clean limit}
The clean limit Eq.~\ref{Eq:IcRn_clean} was fitted (Fig. \ref{IcdFdirty} a) to the data inside the magnetic regime, i.e., $d_F>d_F^\mathrm{dead}$. We obtained an exchange energy $E_\mathrm{ex}=380\:\rm{meV}$ (solid line) assuming $v_F=2.2\cdot10^{5}\:\rm{m/s}$ \cite{PetrovykhNI1998}. However, this fit yields several minima in $j_c$, which can not be seen in our set of data. For completeness we calculated $j_c(d_F)$ for $E_\mathrm{ex}=300$ and $80\:\rm{meV}$.  Again, the agreement with data is bad. Our data has either a larger oscillation period, or the decay length is shorter than for the calculated curve. One may increase $d_F^\mathrm{dead}$ to obtain a slightly better congruence, but at the same time the data set for fitting becomes even smaller.\\ We conclude that the clean limit theory Eq. \ref{Eq:IcRn_clean} can not reproduce our data. Furthermore, the strong decay of $j_c$ inside dead magnetic regime can only be explained by dirty limit condition and after onset of magnetism in $d_F>d_F^\mathrm{dead}$ the transport regime is not expected to modify drastically to the clean limit condition.

\subsubsection*{Dirty limit}
Two fits were done for different data ranges in order to reproduce the experimental data, i.e., Eq.~\ref{Eq:IcRn} was fitted to the total range (regimes I+II+III) and magnetic active range (II+III), respectively. We estimated $\xi_{F1}=0.81\:(0.66)\:\mathrm{nm}$, $\xi_{F2}=1.18 \:(0.53) \:\mathrm{nm}$ and $d_F^\mathrm{dead}=1.2\:(2.26)\:\mathrm{nm}$.\\
For both fits (to I+II+III or II+III) we obtained a short decay length $\xi_{F1} \ll d_F$. The fit to I+II+III yields a rather large $\xi_{F2}$, strong decay ($\xi_{F1} < \xi_{F2}$) and underestimates the decay of $j_c$ inside $\pi$ state, whereas the fit  to II+III has good correlation with data, but yields $\xi_{F1}>\xi_{F2}$,  which -strictly speaking- contradicts the dirty limit theory.\\
Assuming the lowest value for the mean free path, i.e. $\ell=2\:\rm{nm}$ \cite{PetrovykhNI1998}, the values for $\xi_{F1},\xi_{F2}$ gave
\[E_\mathrm{ex}=(\frac{1}{\xi_{F1}}+\frac{1}{\xi_{F2}})^2 \frac{\hbar v_F\ell}{12}=104\:(279)\:\rm{meV}\]
and $\hbar/\tau_m=0.37\:(0.22)\: E_\mathrm{ex}$. Larger values for $\ell$ yield an $E_\mathrm{ex}$ being much larger than the bulk value ($E_\mathrm{ex}\approx310\:\rm{meV}$). However, up to $d_F^\mathrm{dead}=2.26\:\rm{nm}$ no magnetic influence on $j_c$ was observed, and the second dirty limit condition $\ell<\hbar v_F/E_\mathrm{ex}$ was at least valid for very thin $d_F$, when $E_\mathrm{ex}$ was strongly reduced or even vanished. This conclusion is limited by the small range of data, which included the strongly shifted $I_c(H)$ pattern in regime III.

Thus, we state that the samples were dirty within the dead magnetic region. $j_c$ of our $\Ni$-SIFS samples considerably drops inside regime I by a factor of $\sim5$ at the F-layer interface and $\sim e^{d_F^\mathrm{dead}/\xi_{F}^{\textrm{dead}}}\approx20$ inside the dead magnetic layer. If the analysis is limited to the magnetically active part (regimes II+III) indications for dirty limit conditions arise, too, although the set of data is rather small and $\xi_{F1}> \xi_{F2}$ cannot be explained by the theory for Eq. \ref{Eq:IcRn}. For the clean limit condition we should observe some $I_c$ for $d_F>4.4\:\rm{nm}$. It's absence may be caused by the onset of magnetic anisotropy effects in regime III.

\subsection{Comparison with literature}\label{CompLiterature}
\begin{table*}[tb]
\caption{\W{Junction parameters of SFS and SIFS JJs showing magnetic interlayer thickness dependent inversion of the ground state phase. The $j_c$ can not be calculated from Ref.\cite{RobinsonPRB07}. \label{Tab:LiteratureCompare}}}
\begin{tabular}{cccccccccc}
  \hline
  \hline
   magnet & type  & $d_F^\mathrm{dead}$ & $j_c(\pi)$ & $I_cR(\pi)$& $I_cR_n(\pi)$& $T$& dirty/clean& $E_\mathrm{ex}$&Ref.\\
 & & [$\rm{nm}]$ & [$\rm{A/cm^2}$] & $[\rm{\mu V}$] &$[\rm{\mu V}$] &[\rm{K}]&&[\rm{meV}] &\\
  \hline
  $\Ni_{0.6}\Cu_{0.4}$ & SIFS &3.09&5 & 400&28&2.11 &dirty&99& \cite{WeidesHighQualityJJ}\\
  $\Ni_{0.53}\Cu_{0.47}$ & SFS &4.3&1000 & 0.15&-&4.2 &dirty&73& \cite{OboznovRyazanov06IcdF}\\
  $\Ni$ & SIFS &2.26&3.4& 7.3&3.7&4.2 &dirty&279& this work\\
  $\Ni$ & SFS &-&1000&0.2 &-&4.2 &clean&200& \cite{Blum04NanoGiga,ShelukhinSFS06}\\
  $\Ni$ & SFS &1.3&-&100& -&4.2 &clean&80& \cite{RobinsonPRB07}\\
  $\Co$ & SFS &0.8&-&60 &-&4.2 &clean&309& \cite{RobinsonPRB07}\\
  $\Fe$ & SFS &1.1&-&125&- &4.2 &clean&256& \cite{RobinsonPRB07}\\
  $\Ni_{0.8}\Fe_{0.2}$ & SFS &0.5&-&80&- &4.2 &clean&201& \cite{RobinsonPRB07}\\
  $\Pd_{88}\Ni_{12}$ & SIFS &-&0.036&-&18 &1.5 &dirty&201& \cite{Kontos02Negativecoupling}\\
   \hline
  \hline
\end{tabular}
\end{table*}

Compared to SFS-JJs with $\Ni$ as an interlayer \cite{Blum:2002:IcOscillations,Blum04NanoGiga,ShelukhinSFS06,RobinsonPRB07}, where multiple $I_c(d_F)$ oscillations were possibly observed, we can determine just one oscillation in our SIFS samples due to both the dirty transport regime and the onset of anisotropy. \W{Our SIFS-JJs, made by multilayer process, optical lithography and ion etching, are good integratable into standard digital logics like RSFQ logic. The FIB-patterned SFS-JJs \cite{RobinsonPRB07} are not suitable for integration into complex circuits, and the SFS-stacks made by Ref.\cite{Blum:2002:IcOscillations,Blum04NanoGiga,ShelukhinSFS06} were not fabricated in one run, and some degree of irreproducibility during deposition or patterning may have occurred, leading to an increased spread of data. We regard our in-situ multilayer deposition as being superior, especially regarding the quality of interfaces.} Over and above for all SFS-JJs just a few data points were obtained. For example, the oscillation period determined in early work \cite{Blum:2002:IcOscillations} was later corrected \cite{Blum04NanoGiga,ShelukhinSFS06} by samples with some closer spacing of $d_F$.\\
On one hand, our more sophisticated stacks -containing an $\Al_2\O_3$ tunnel barrier and a thin structural $\Cu$ layer- are subject to stronger scattering at the interfaces and therefore more likely to have a $j_c$ below the measurement resolution. On the other hand \W{our samples have very smooth lower SI interface, which is secured by the observation of tunneling. The local current density depends exponentially on tunnel barrier thickness, and a variation in $\Al_2\O_3$ thickness would provoke pinholes and magnetic-field independent IV characteristics. The interface roughness is much smaller than the tunnel barrier thickness $\approx1\:\rm{nm}$.} Our larger density of data than in Ref. \cite{Blum:2002:IcOscillations,Blum04NanoGiga,ShelukhinSFS06,RobinsonPRB07} yields more information both on i) the variation of $j_c$ for same $d_F$ (very low due to same run deposition and patterning) and ii) the transport close to the onset of magnetism in F-layer.\\
In Tab.\ref{Tab:LiteratureCompare} we give an overview on the current status on $\pi$ coupled JJs being SFS or SIFS type. The rather high $j_c$ of SFS junctions is achieved by low interface scattering, and therefore they have a considerably lower junction resistance as the SIFS-type junctions.\\The large amplitude of the subgap-current, which depends on $d_F$ and $T$, complicates the determination of the junction resistance. Usually for SIS junctions the normal state, ohmic resistance $R_n$, measured beyond $V>2\Delta/e$ is considered for the quality factor. Current-biased SIFS-JJs jump from the Meissner-state to a voltage $V<2\Delta/e$ which depends on the subgap-resistance $R$. However, for SFS-type junctions $R_n=R$ is taken as the constant resistance branch for $I>I_c$ and at $V\ll\Delta/e$. For implementation of SIFS-JJs the $I_cR$ product is only relevant, too. In first work on SIFS-JJs \cite{Kontos02Negativecoupling} $I_cR_n(d_F)$ instead of $I_c(d_F)$ was used to avoid data scattering due to variations of $R_n$. However, in this paper $R_n$ is constant for all JJs within an experimental error of about 5\%, thus we plot $j_c(d_F)$.\\The exchange energy for $\Ni$ and its $\Cu$ alloys is consistently ranging between $73\:\rm{meV}$ \cite{OboznovRyazanov06IcdF}  up to $279\:\rm{meV}$ (this work), except the considerable lower value  in Ref. \cite{RobinsonPRB07} for pure $\Ni$ ($80\:\rm{meV}$). The Ref. \cite{RobinsonPRB07} does not provide information about the critical current density $j_c$, as the significant variation of junction area is overcome by considering just $I_cR$. One may speculate if these focused ion beam etched JJs resemble more S(FN)S-type JJs \cite{Kypriyanov_SFNS_07}, where the interlayer consists of a ferromagnetic core being surrounded by a normal metal. This may explain the considerable lower exchange energy ($80\:\rm{meV}$, like $\Ni_{0.53}\Cu_{0.47}$ alloy in Ref. \cite{OboznovRyazanov06IcdF}) and the significant enhanced period of $I_c$ oscillations ($\sim4\:\rm{nm}$).\\SIFS junctions with $\Ni$ (this work) and $\Ni\Cu$ \cite{WeidesHighQualityJJ} as magnetic interlayer have similar $j_c$'s, but $R_n$ and $I_cR_n$ of the $\Ni\Cu$ based SIFS JJs are 5 times larger. The scattering probability in $\Ni$ and therefore the excess current are increased by this factor compared to $\Ni\Cu$ alloys. Furthermore, the thicker $\Al_2\O_3$ barrier in $\Ni\Cu$ SIFS-JJs reduces the subgap current created by microshorts in the tunnel barrier, and provides a larger subgap resistance $R$.\\We would like to point out that the use of $\Ni$ for $\pi$ JJs has several disadvantages: i) strong scattering of supercurrent, ii) large dead magnetic layer where $j_c$ is already reduced by a factor of $40$ and iii) anisotropy effects for $d_F>3.8\:\rm{nm}$. SIFS-JJs with $\Co$ and $\Fe$, having large atomic magnetic moments, may display anisotropy effects even for thinner magnetic thicknesses.\\Based on the data of Tab. \ref{Tab:LiteratureCompare} the magnetically diluted $\Pd$ alloys may be an alternative for $\pi$ SIFS JJs with high $j_c$, as the absolute drop of $j_c$ due to F-layer is only about a factor of 20\textrm{-}77 \cite{Kontos02Negativecoupling}.

\section{Conclusions}

In summary, \W{for the first time SIFS-JJs with a strong magnetic interlayer,} i.e. $\Ni$ and an $\Al_2\O_3$ tunnel barrier were studied. \W{Our samples had a large density of data points, ranging from the magnetically dead regime towards very thick $d_F$ layers, being larger than in previous work on SFS JJs with elemental magnets \cite{Blum:2002:IcOscillations,Blum04NanoGiga,ShelukhinSFS06,RobinsonPRB07}, where the data spacing was of the same order of magnitude as the phase oscillation lengths. Thus, we show the first results that allow to find the oscillation period for the elemental magnets reliably.} The insertion of F-layer leads to additional interface scattering compared to non-magnetic junctions and inside the dead magnetic layer $j_c$ drops exponentially. \W{The dead magnetic layer thickness $d_F^\mathrm{dead}$ has been determined directly from transport measurements.} The critical current $I_c$ changes its sign as a function of the F-layer thickness $d_F$, exhibiting regions with $0$ and $\pi$ ground states. For $d_F$ near the $0$ to $\pi$ crossover the ground state can be controlled by changing the temperature. This is the first observation of a temperature-induced phase change using a strong magnet. For certain thicknesses the junctions show magnetic anisotropy effects, leading to a distortion of their $I_c(H)$ pattern. Overall, the transport regime is dirty, although locally inside the magnetically active F-layer regime a deviation from the strict dirty limit theory appears.

\section*{Acknowledgement}
The authors thank H. Kohlstedt, D. Sprungmann, E. Goldobin, D. Koelle, R. Kleiner, N. Pugach, M. Yu. Kupriyanov and A. Palevski for stimulating discussions. J.P. is supported by Studienstiftung des Deutschen Volkes, V.S.S. and V.V.R by the MOST-RFBR project 06-02-72025 and M.W. by DFG project WE 4359/1-1.


\end{document}